\documentclass[11pt,a4paper]{article}
\pdfoutput=1
\usepackage{jheppub}

\usepackage{amsmath}
\usepackage{amsfonts}
\usepackage{amssymb}

\usepackage{float}

\title{Simulations of Cold Electroweak Baryogenesis: Dependence on the source of CP-violation}

	\author[b]{Zong-Gang Mou,}
	\author[a]{Paul M. Saffin,}
	\author[b]{Anders Tranberg}

	\affiliation[a]{School of Physics and Astronomy, University Park, University of Nottingham,\\ Nottingham NG7 2RD, United Kingdom}
	\affiliation[b]{Faculty of Science and Technology, University of Stavanger, 4036 Stavanger, Norway}

	\emailAdd{zonggang.mou@uis.no}
	\emailAdd{paul.saffin@nottingham.ac.uk}
	\emailAdd{anders.tranberg@uis.no}
	

\abstract{We compute the baryon asymmetry created in a tachyonic electroweak symmetry breaking transition, focusing on the dependence on the source of effective CP-violation. Earlier simulations of Cold Electroweak Baryogenesis have almost exclusively considered a very specific CP-violating term explicitly biasing Chern-Simons number. We compare four different dimension six, scalar-gauge CP-violating terms, involving both the Higgs field and another dynamical scalar coupled to SU(2) or U(1) gauge fields. We find that for sensible values of parameters, all implementations can generate a baryon asymmetry consistent with observations, showing that baryogenesis is a generic outcome of a fast tachyonic electroweak transition. }

\begin{document}

\maketitle

\section{Introduction}
\label{sec:Intro}

Cold Electroweak Baryogenesis attempts to explain the observed baryon asymmetry in the Universe by postulating that the process of electroweak symmetry breaking was a cold spinodal transition \cite{Krauss:1999ng,GarciaBellido:1999sv,Copeland:2001qw,Tranberg:2003gi}. This is possible if the Higgs field $\phi$ is coupled to another field, whose dynamics triggers symmetry breaking only after the Universe has cooled below the electroweak scale \cite{Copeland:2001qw,vanTent:2004rc,Enqvist:2010fd,Konstandin:2011ds}.  In such a cold transition, a baryon asymmetry is created in the presence of CP-violation, as the out-of-equilibrium conditions required for successful baryogenesis are provided by the exponentially growing IR modes of the spinodal (Higgs) field. C and P violation follow from the electroweak sector of the Standard Model. As for traditional (hot) electroweak baryogenesis, the CP-violation arising from the Standard Model CKM matrix is insufficient \cite{Gavela:1994dt,Gavela:1994ds,Brauner:2011vb}. Sources of CP-violation beyond the Standard Model must therefore be part of the scenario.

In a series of recent papers \cite{Mou:2017atl,Mou:2017zwe,Mou:2017xbo}, using classical lattice field theory simulations we have studied the effect of relaxing a sequence of assumptions of the original work \cite{GarciaBellido:2003wd,Tranberg:2003gi,Tranberg:2006ip,Tranberg:2006dg}. This includes the dependence on the speed of the spinodal transition \cite{Mou:2017xbo},  the impact of U(1) hypercharge gauge fields on the asymmetry \cite{Mou:2017zwe}, and the effect of replacing a ``by-hand" mass-flip of the Higgs field by a portal coupling to a new dynamical field $\sigma$ \cite{Mou:2017atl}.

In the present work, we relax one final assumption, namely the introduction of CP violation through one specific dimension-6 term
\begin{eqnarray}
\label{eq:CP2p}
S_{\rm 2,\phi}=\frac{3\delta_{2,\phi}g^2}{16\pi^2m_{\rm W}^2}\int dt\,d^3x \,\phi^\dagger\phi \textrm{Tr}\,W^{\mu\nu}\tilde{W}_{\mu\nu},
\end{eqnarray}
with $W^{\mu\nu}$ the field strength tensor of the SU(2) gauge field and $\tilde{W}_{\mu\nu}=\frac{1}{2}\epsilon_{\mu\nu\rho\sigma}W^{\rho\sigma}$. The dimensionless constant $\delta_{2,\phi}$ is a measure of the magnitude of CP-violation, and could in principle be derived from matching this effective term to some underlying theory. $\phi^\dagger\phi$ is manifestly C and P even, and $W\tilde{W}$ is C even, but P odd.
The common feature of all electroweak baryogenesis scenarios is that the baryon asymmetry arises from generating a non-zero value of Chern-Simons number
\begin{eqnarray}
\label{eq:ncs2}
N_{\rm cs,SU(2)}(t)-N_{\rm cs,SU(2)}(0) = \frac{g^2}{16\pi^2}\int_0^t dt\, d^3x\, \textrm{Tr}\, W^{\mu\nu}\tilde{W}_{\mu\nu},
\end{eqnarray}
since baryon number then changes according to the chiral anomaly
\begin{eqnarray}
\label{eq:anomaly}
3[N_{\rm cs,SU(2)}(t)-N_{\rm cs, SU(2)}(0)]=B(t)-B(0).
\end{eqnarray}
It is clear that the term (\ref{eq:CP2p}) has a very special standing, in that by partial integration and assuming that $\phi$ is approximately constant in space, one gets
\begin{eqnarray}
\label{eq:bias}
S_{\rm 2,\phi}\simeq -\frac{3\delta_{2,\phi}}{m_{\rm w}^2} \int dt\, \partial_0(\phi^\dagger\phi)N_{\rm cs,SU(2)}.
\end{eqnarray}
As soon as $\phi$ changes in time, an effective bias is introduced precisely for the Chern-Simons number which then generates a baryon asymmetry.

In a more generic model, one would expect CP-violation to be present in the system, but not as an explicit bias in this way. More likely, during the transition CP-violation forces the complete set of fields to favour CP-violating configurations, and in such a background, Chern-Simons number is effectively biased to a non-zero expectation value. 

Modelling the Standard Model through an effective bosonic theory including only the Higgs field $\phi$ and SU(2) gauge field $W_\mu$, Eq. (\ref{eq:CP2p}) is the natural lowest order CP-violating term (although not the only one, see \cite{Brauner:2011vb}). But including also U(1) gauge fields and a symmetry-triggering scalar $\sigma$, as necessary for achieving a cold tachyonic transition (see below), other possibilities arise, including
\begin{eqnarray}
\label{eq:all3}
S_{\rm 2, \sigma}&=&\frac{3\delta_{2,\sigma}g^2}{16\pi m_{\rm W}^2}\int dt\, d^3x\, \xi^2\sigma^2 \, \textrm{Tr}\,W^{\mu\nu}\tilde{W}_{\mu\nu},\\
S_{\rm 1,\phi}&=&\frac{3\delta_{1,\phi}(g')^2}{32\pi m_{\rm W}^2}\int dt\, d^3x\, \phi^\dagger\phi \, B^{\mu\nu}\tilde{B}_{\mu\nu},\\
S_{\rm 1,\sigma}&=&\frac{3\delta_{1,\sigma}(g')^2}{32\pi m_{\rm W}^2}\int dt\, d^3x\,\xi^2\sigma^2 \, B^{\mu\nu}\tilde{B}_{\mu\nu},
\end{eqnarray}
with $B_{\mu\nu}$ the U(1) (hypercharge) gauge field strength. New parameters $\delta_{2,\sigma}$, $\delta_{1,\phi}$, $\delta_{1,\sigma}$, are introduced representing the magnitude of CP-violation. $\xi$ is a dimensionless portal coupling to be defined below.
Whereas the first of these terms again biases $N_{\rm cs, SU(2)}$ (a {\it primary} bias, in our terminology), the next two bias another CP-odd observable (the U(1)-Chern-Simons number) 
\begin{eqnarray}
\label{eq:ncs1}
N_{\rm cs,U(1)}(t)-N_{\rm cs,U(1)}(0) = \frac{(g')^2}{32\pi^2}\int_0^t dt\, d^3x\,  B^{\mu\nu}\tilde{B}_{\mu\nu},
\end{eqnarray}
which then through the field dynamics potentially biases $N_{\rm cs,SU(2)}$ (a {\it secondary} bias).

Establishing whether, and under what conditions, such a secondary bias is able to generate sufficient baryon asymmetry is the purpose of this work. Clearly, secondary bias is the most generic source of CP-violation and, if successful, opens up new paths of model building for this baryogenesis scenario. A combination of the two was considered in \cite{Tranberg:2012jp,Tranberg:2012qu,Mou:2015aia} for the 2-Higgs doublet model where, instead of (\ref{eq:CP2p}), the authors considered
\begin{eqnarray}
S_{\rm 2hdm}= \frac{3\delta_{\rm 2hdm}g^2}{16\pi m_{\rm W}^2}\int dt\,d^3x\,(\phi^\dagger_1\phi_2-\phi^\dagger_2\phi_1)\textrm{Tr }W^{\mu\nu}\tilde{W}_{\mu\nu}.
\end{eqnarray}
This works as a primary bias, breaks both C and P, but conserves CP. In addition, it is then necessary to include C-violation in the 2-Higgs potential, effectively to bias the combination $\phi^\dagger_1\phi_2-\phi^\dagger_2\phi_1$ to be nonzero. This was seen to generate a large enough baryon asymmetry to match observations \cite{Tranberg:2012jp,Tranberg:2012qu}.

In the following section \ref{sec:model}, we present our model: the bosonic part of the electroweak sector of the Standard Model, coupled to a singlet scalar. We further discuss the four different CP-violating terms that we will consider, and present some discussion about CP-odd observables and how they are related. In section \ref{sec:cewbag} we give a brief overview of Cold Electroweak Baryogenesis and show a few examples of the behaviour of the observables. In section \ref{sec:results} we then compare the asymmetries resulting from each of the four CP-violating terms and when some of them are combined. We also comment on the effect of a constant (in time and space) bias of $N_{\rm cs, SU(2)}$, and lattice discretization effects. We conclude in section \ref{sec:conclusion}.

\section{Model}
\label{sec:model}

Building on the work of \cite{Mou:2017atl}, we consider the bosonic part of the Standard Model electroweak sector, extended by a singlet scalar $\sigma$ coupled to the Higgs field $\phi$. The action reads:
\begin{eqnarray}
\label{eq:S_EW}
S= \int dt\, d^3x\Bigg[ &-\frac{1}{2}\textrm{Tr}\,W^{\mu\nu}W_{\mu\nu} -\frac{1}{4} B^{\mu\nu}B_{\mu\nu}
- (D_\mu\phi)^\dagger D^\mu\phi +\mu^2\phi^\dagger\phi-\lambda\left(\phi^\dagger\phi\right)^2-\frac{\mu^4}{4\lambda}
\nonumber \\ 
 &-\frac{1}{2}\partial_\mu\sigma\partial^\mu\sigma-\frac{m^2}{2}\sigma^2 - \frac{1}{2}\xi^2 \sigma^2 \phi^\dagger\phi
\Bigg]+ S_{\rm CP},
\end{eqnarray}
where for the SU(2) gauge field, we have $W_{\mu\nu} = \partial_\mu W_\nu - \partial_\nu W_\mu -ig[W_\mu, W_\nu]$,
$W_\mu= W_\mu^a\sigma^a/2$ with $\sigma^a$ the Pauli matrices, and similarly for the U(1) hypercharge field $B_{\mu\nu} = \partial_\mu B_\nu - \partial_\nu B_\mu$. The covariant derivative is given by
\begin{eqnarray}
D_\mu\phi = \left(\partial_\mu -i Yg' B_\mu-i gW_\mu\right)\phi,
\end{eqnarray}
with $Y=-1/2$ for the Higgs field.

We have explicitly put in the Higgs vacuum expectation value $v=246$ GeV, the Higgs self-coupling $\lambda = \mu^2/v^2=m_H^2/(2v^2)\simeq 0.13$, and the gauge couplings $g=0.65$ and $g'=0.35$. This corresponds to $m_H=125$ GeV, $m_W=80$ GeV, and $m_Z=91$ GeV. In addition, we have the free parameters of the $\sigma$-$\phi$ potential, $m^2$ and $\xi$. We have chosen a very simple potential form, ignoring cubic and quartic $\sigma$ self-interactions and the cubic portal coupling. This is just for simplification and to match \cite{Mou:2017atl}. Engineering the $\sigma$-potential to have more features (non-zero expectation values in the vacuum, away from the vacuum) may have implications for the baryon asymmetry. 

We will stick to the quadratic form indicated in (\ref{eq:S_EW}). In the language of \cite{Mou:2017atl}, we will consider a fast ($m_H/m=4$ and $\xi=2.04$) and slow ($m_H/m=32$ and $\xi=0.254$) quench at $n=8$, where $n$ indicates the total energy in the system through
\begin{eqnarray}
V_{\rm tot}= V_0 \left(1+\frac{1}{n^2}\right)=\frac{\mu^4}{4\lambda} \left(1+\frac{1}{n^2}\right).
\end{eqnarray}
For $n=8$, the energy initially stored in the non-zero $\sigma$ field is therefore negligible (about 1\%) compared to $V_0$, the potential energy density from the Higgs potential itself at $\phi=0$, $\sigma=0$. For more details of this point, we refer the reader to \cite{Mou:2017atl}.

As advertised in the introduction, we will consider four different effective bosonic dimension-6 terms playing the role of $S_{\rm CP}$.  In previous work, we found that a baryon asymmetry consistent with observations corresponds to $\delta_{2,\phi}\simeq 10^{-5}$, with some dependence on the speed of the symmetry breaking quench \cite{Mou:2017atl}.

The full Standard Model includes all the fermions as well, with CP-violation encoded in the CKM-matrix. It is tempting to expect that when integrating these out, CP-violation would be recovered as terms of the form (\ref{eq:CP2p}), (\ref{eq:all3}). This is true in terms of the field content, but the structure of the effective terms is rather more complex \cite{Brauner:2011vb}. Also, the magnitude of the coefficients $\delta_{i,j}$ is much too small to be responsible for baryogenesis, unless the effective temperature during the transition is less that 1 GeV \cite{Brauner:2011vb}, which does not seem to be the case \cite{Mou:2013kca}.

So for our purposes, although we do expect that such effective terms arise from integrating some heavier degrees of freedom, they are just generic representatives of CP-violation providing primary and secondary bias. 

\subsection{Observables}
\label{sec:obs}

As we have no fermions explicitly in the system, we rely on the chiral anomaly relation (\ref{eq:anomaly}) to infer the baryon asymmetry. But in fact, in the presence of U(1) gauge fields in addition to the SU(2) gauge fields, the full chiral anomaly is the sum of two contributions 
\begin{eqnarray}
\label{eq:anomaly2}
B(t)-B(0) = 3\left[N_{\rm cs,SU(2)}(t)-N_{\rm cs,SU(2)}(0)\right]-3\left[N_{\rm cs, U(1)}(t)-N_{\rm cs, U(1)}(0)\right].
\end{eqnarray}
Usually, this complication is ignored, as one is interested in permanent changes of the Chern-Simon number. For the SU(2) gauge theory, the vacuum structure consists of a series of gauge equivalent vacua with integer Chern-Simons number. Hence, going from one minimum to the next produces net baryon number, and this asymmetry can remain at late times and low temperatures. The vacuum structure for the U(1) gauge field is trivial, with a single vacuum at $N_{\rm cs,U(1)}=0$. This means that although during the process, U(1) Chern-Simons number may be biased to one side, ultimately it will relax back to zero, restoring the simple form (\ref{eq:anomaly}).

As a further proxy for the baryon asymmetry, we note that the Higgs field winding number
\begin{eqnarray}
\label{eq:nw}
N_{\rm w}=\frac{1}{24\pi^2}\int d^3 x \epsilon^{ijk}\textrm{Tr}[(U^\dagger\partial_i U )(U^\dagger\partial_jU )(U^\dagger\partial_k U)],
\end{eqnarray}
with $U(x)=(i\tau_2\phi^*,\phi)/\phi^\dagger\phi$, in a ``pure-gauge" vacuum obeys
\begin{eqnarray}
N_{\rm w}= N_{\rm cs,SU(2)}.
\end{eqnarray}
This follows from the minimization of the covariant derivative, when $B_{\mu}=0$. But more generally, we have the relation
\begin{eqnarray}
\label{eq:puregauge}
N_{\rm w}\simeq N_{\rm cs,SU(2)}-N_{\rm cs,U(1)},
\end{eqnarray}
a relation we will confirm numerically below. Because $N_{\rm w}$ is integer (up to lattice artefacts) and therefore a much less noisy numerical observable, we will make the identification at late times
\begin{eqnarray}
B(t)-B(0) = 3[N_{\rm w}(t)-N_{\rm w}(0)].
\end{eqnarray}

In our simulations we will average the observables over an initially CP-symmetric ensemble of field realisations, initialised to reproduce the correlation functions of the quantum vacuum \cite{GarciaBellido:2002aj,Smit:2002yg}. The dynamics themselves follow the classical equations of motion, as derived from the full lagrangian. The detailed numerical lattice implementation may be found elsewhere \cite{Tranberg:2003gi}.

To track the progress of the transition, we will often plot the average Higgs field
\begin{eqnarray}
\langle\phi^2\rangle = \frac{1}{V}\int d^3x\, \phi^\dagger\phi(x),
\end{eqnarray}
and $\sigma$ field 
\begin{eqnarray}
\langle\sigma\rangle = \frac{1}{V}\int d^3x\, \sigma(x),
\end{eqnarray}
also averaged over the ensemble.

\section{Cold Electroweak Baryogenesis}
\label{sec:cewbag}

Detailed expositions of many aspects of the Cold Electroweak Baryogenesis scenario is available in the literature \cite{Tranberg:2003gi,vanderMeulen:2005sp}. In brief, the non-Standard Model degree of freedom $\sigma$ is assumed to start out at a value $\sigma_i>\sigma_c=\mu/\xi$, and to roll down its potential to $\sigma=0$. In doing so, the mass parameter of the Higgs field changes sign, with
\begin{eqnarray}
\mu^2_{\rm eff}(t) = \xi^2\sigma^2(t)-\mu^2.
\end{eqnarray}
We will take $\sigma_i=\sqrt{2}\sigma_c$, in such a way that $\mu^2_{\rm eff}$ goes from $+\mu^2$ initially to $-\mu^2$ asymptotically at late times. Although the exact trajectory by which this happens will depend on the parameters of the model, ultimately this will result in electroweak symmetry breaking. 

While $\mu_{\rm eff}^2(t)<0$, momentum modes of the Higgs field with $k^2+\mu^2_{\rm eff}(t)<0$ grow exponentially, a process known as spinodal transition or tachyonic preheating. The result is that the energy in the initial Higgs potential is transferred to particles in the IR ($k<\mu$) of the spectrum. The instability itself, but also the subsequent redistribution of energy into the UV, are strongly out of equilibrium processes, suitable for generating a baryon asymmetry.

The speed of the transition may be expressed as 
\begin{eqnarray}
u= -\frac{1}{2\mu^3}\frac{d\mu^2_{\rm eff}}{dt}|_{\mu^2_{\rm eff}=0}\equiv \frac{1}{\mu\tau_q},
\end{eqnarray}
with $\tau_q$ a characteristic quench time. We found in \cite{Mou:2017xbo} for the exact same model considered here the relation $\tau_q\simeq 1.3\, m^{-1}$, and so from now on, we will express the quench time in terms of the dimensionless ratio $m_H/m\simeq 0.8\,m_H\tau_q\simeq 1.1/u$.
The maximum asymmetry occurs for quench times $m_H/m\simeq 30$, whereas very fast quenches with $m_H/m\simeq 0$, most favoured by model-building, give an asymmetry of the opposite sign and a factor of 3-4 smaller in magnitude \cite{Mou:2017atl,Mou:2017xbo}.

A more detailed analysis of the field configurations arising in such a transition shows, that an asymmetry is generated first as the Chern-Simons number is biased to one side by CP-violation, and that subsequently the Higgs winding number changes to accommodate this. And that this happens most readily when there are many points in space with $\phi^\dagger\phi(x)\simeq 0$ \cite{vanderMeulen:2005sp}.

\begin{figure}
\begin{center}
\includegraphics[width=12cm,angle=0]{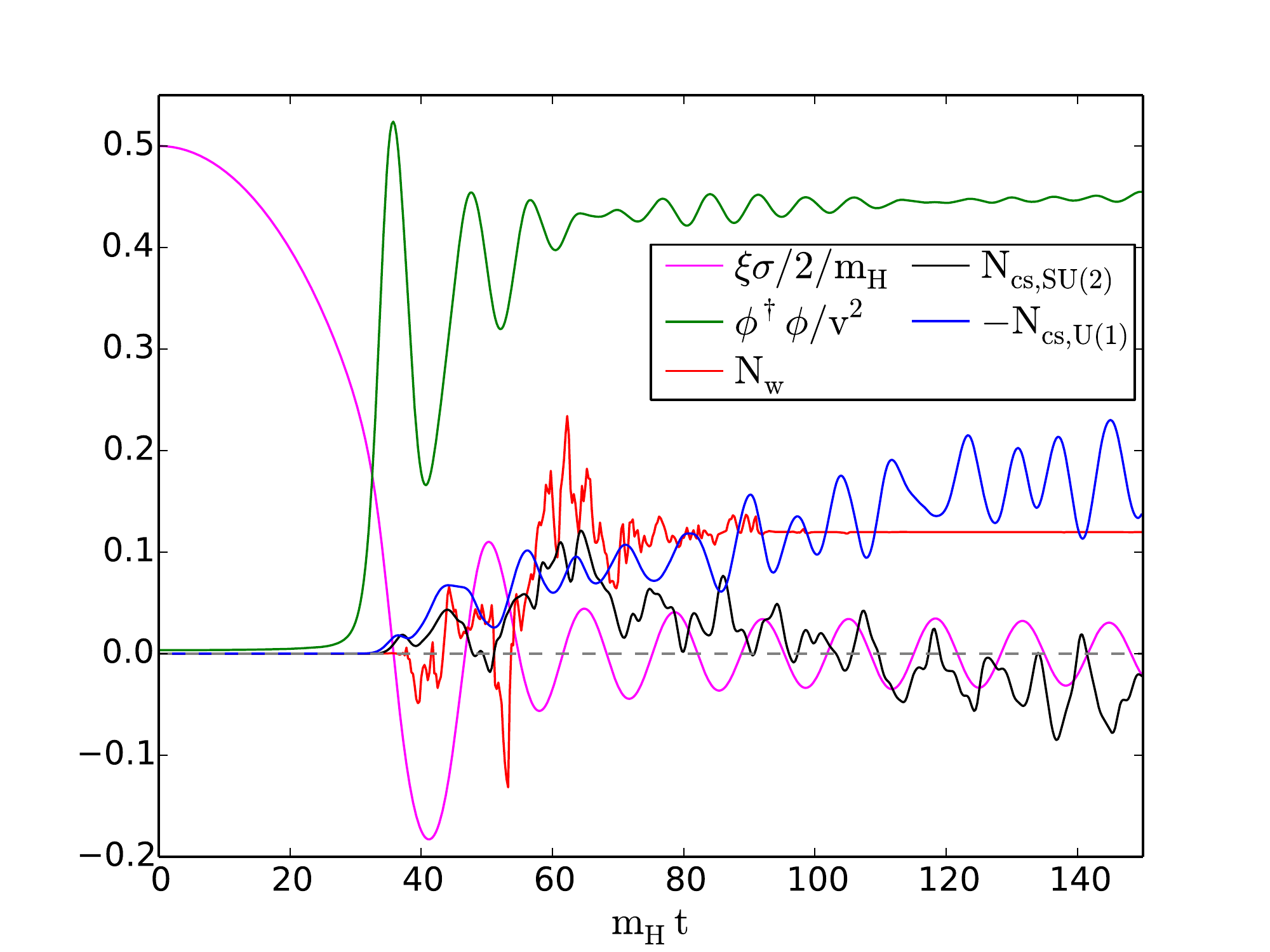}
\caption{The Higgs and $\sigma$ fields and the CP-odd observables in a typical simulation, averaged over an ensemble of 50 CP-conjugate pairs.}
\label{fig:cewbagex}
\end{center}
\end{figure}

In Fig. \ref{fig:cewbagex}, we show the basic observables during the transition, averaged over the ensemble of initial conditions. The quench time is chosen to be $m_H/m=32$, and so until $m_Ht\simeq 25$, the Higgs field is stable at $\phi^2=0$. Then as the effective mass parameter $\mu^2_{\rm eff}$ becomes negative, the Higgs field grows from zero to near the vacuum expectation value $\phi^2/v^2=1/2$, after which it oscillates with a decreasing amplitude. 

Meanwhile, the SU(2) Chern-Simons number (\ref{eq:ncs2}), Higgs winding number (\ref{eq:nw}) and U(1) Chern-Simons number (\ref{eq:ncs1}) deviate from zero average in a complicated way under the influence of CP-violation (here, (\ref{eq:CP2p}). The Chern-Simons number moves first, but for $N_{\rm w}$, most of the motion happens near $m_Ht=40$ and $55$, when the Higgs field is at a minimum in its oscillation.  This is when many local zeros of the Higgs field are present. 

By time $m_Ht=90$, the Higgs field has settled, and the Higgs winding number is completely frozen in. In principle, equilibrium Sphaleron processes could trigger a change in winding and Chern-Simons number, but at an effective temperature way below the critical temperature of the electroweak phase transitions (about $40$ GeV compared to $T_c=160$ GeV \cite{DOnofrio:2014rug}) this is completely negligible. 

\begin{figure}
\begin{center}
\includegraphics[width=7cm,angle=0]{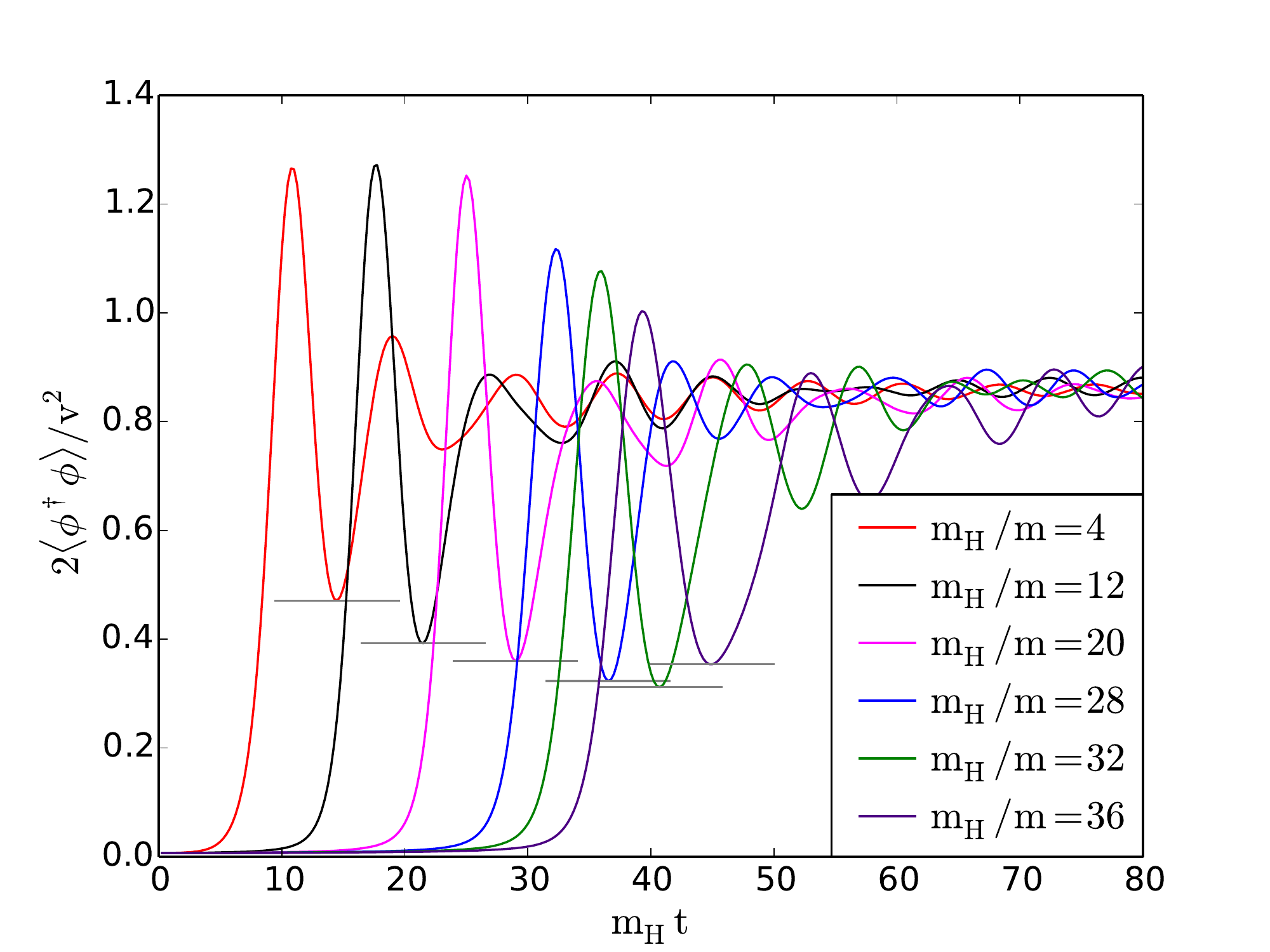}
\includegraphics[width=7cm,angle=0]{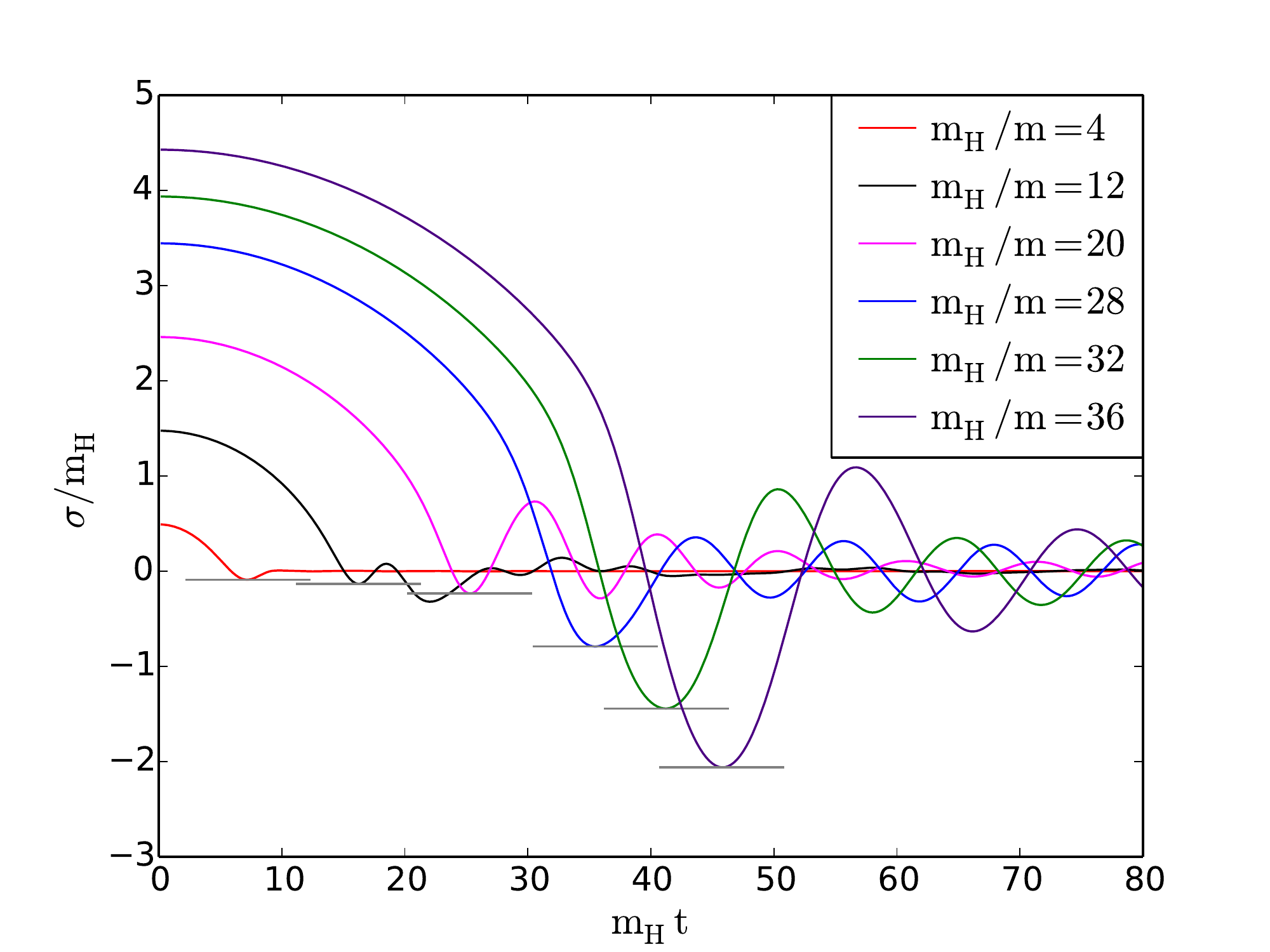}
\caption{The Higgs (left) and singlet (right) fields early in the transition for a range of transition speeds. }
\label{fig:zeros}
\end{center}
\end{figure}

It is a generic feature that the largest asymmetry is created for parameter values giving the largest number of Higgs zeros. In Fig. \ref{fig:zeros}, we show the average Higgs field squared (left) and the singlet field (right) for a number of transition speeds. We see that the Higgs field increases as the transition is triggered, but then oscillates back to a minimum.The value of this minimum decreases with increasing quench time up to $m_H/m=32$, after which it increases again. 

\begin{figure}
\begin{center}
\includegraphics[width=7cm,angle=0]{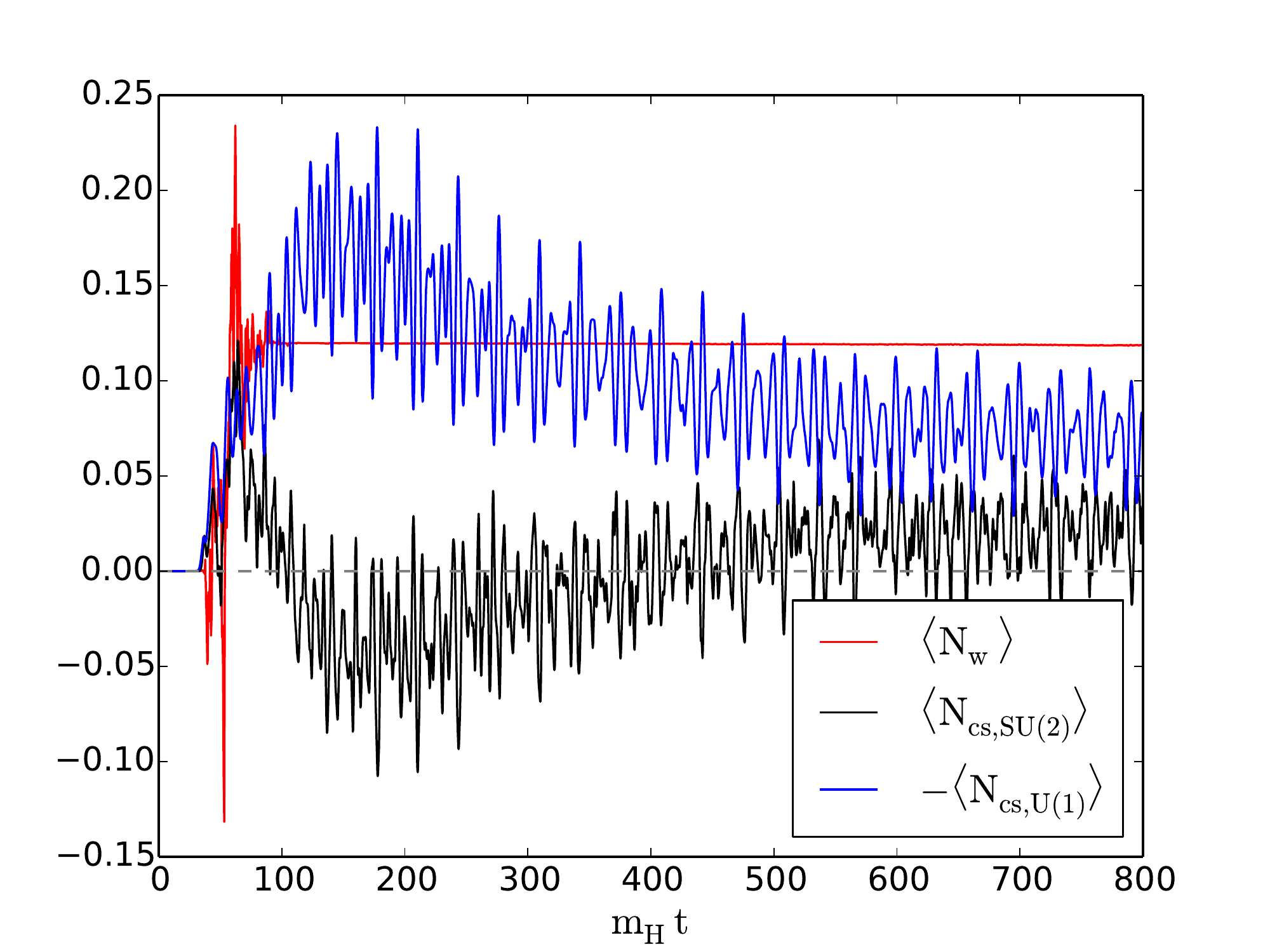}
\includegraphics[width=7cm,angle=0]{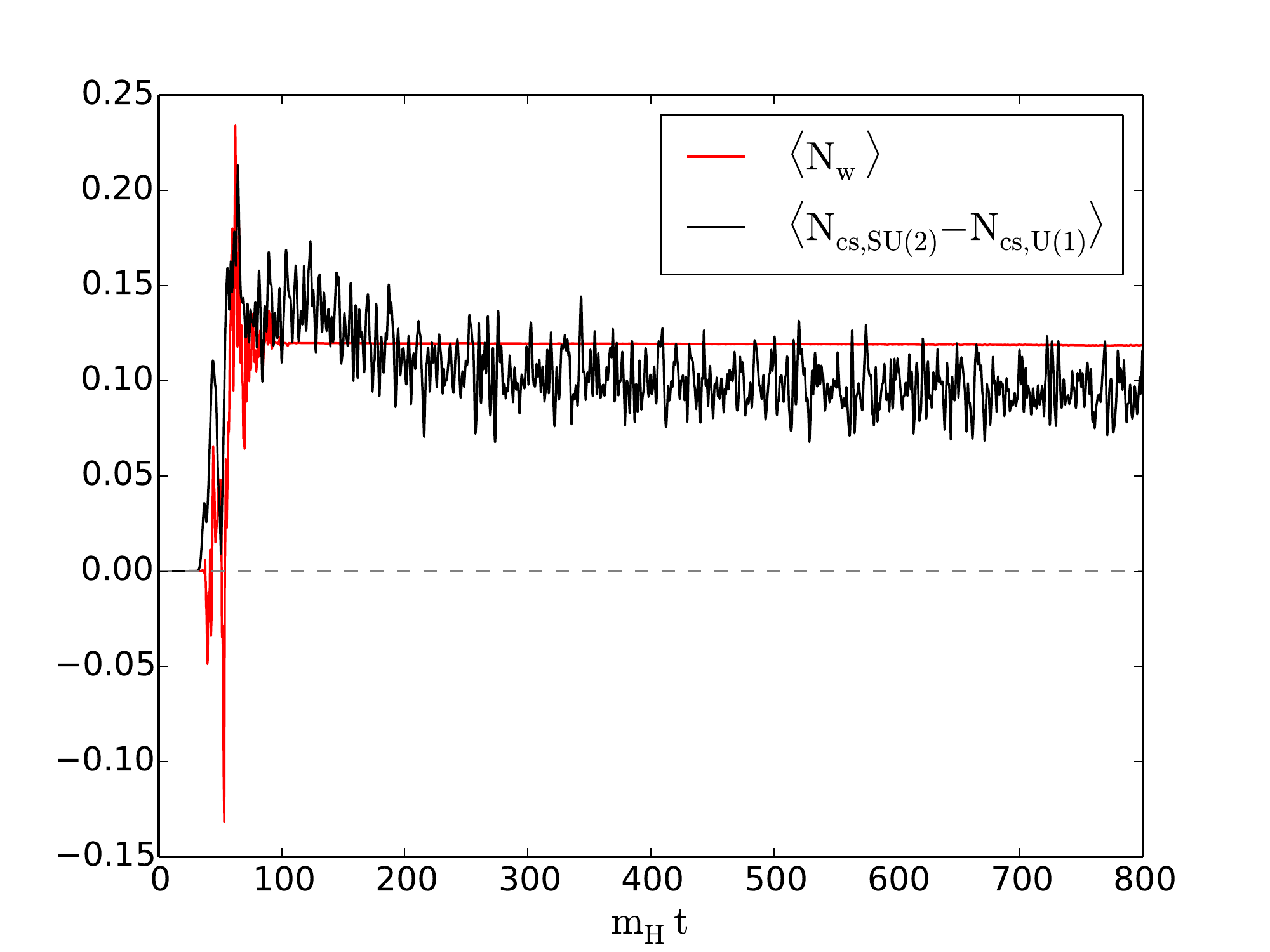}
\caption{The CP-odd observables in a typical simulation, with $N_{\rm cs,SU(2)}$ and $N_{\rm cs,U(1)}$ separately (left) and added up (right).}
\label{fig:sumNcs}
\end{center}
\end{figure}

Returning to Fig. \ref{fig:cewbagex}, we find that Chern-Simons numbers individually do not seem to match the winding number very well, as would be expected for a pure-gauge field configuration. In Fig. \ref{fig:sumNcs}, we show the same observables in the same simulation, but for much longer time. In the left-hand plot, we see the two Chern-Simons numbers separately, whereas in the right-hand plot, we have added them up as in (\ref{eq:anomaly2}). We see that the relation (\ref{eq:puregauge}) applies. We have checked that for very long times, $N_{\rm cs, U(1)}$ indeed goes to zero, so that $N_{\rm w}=N_{\rm cs,SU(2)}$ is restored as a simple proxy for the baryon asymmetry. In what follows, we will use the value of $N_{\rm w}$ at the end of the simulation as our primary observable.

\section{Comparing sources of CP-violation}
\label{sec:results}

The numerical procedure is then for each of the four CP-violating terms to vary the coefficients $\delta_{i,j}$ for the two different quench speeds $m_H/m=4$ (fast) and $m_H/m=32$ (slow), but otherwise keeping parameters fixed. The lattice size $64^3$ and lattice spacing $am_H=0.375$ are kept fixed unless explicitly stated otherwise. The ensemble members are randomly generated, and we use different random seeds for different simulations. The ensembles each consist of 400 CP-conjugate pairs. For each pair of CP-conjugate configurations, we record whether the final values of $N_{\rm w}$ cancel to zero (one is minus the other). If not, we say that the pair has performed a ``flip". Flipped pairs usually add up to $\pm 1$, but instances of $\pm 2$ and $3$ were observed. Statistics and errors are based on the frequency of flips.

\subsection{SU(2)-type CP-violation}

\begin{figure}
\begin{center}
\includegraphics[width=7cm,angle=0]{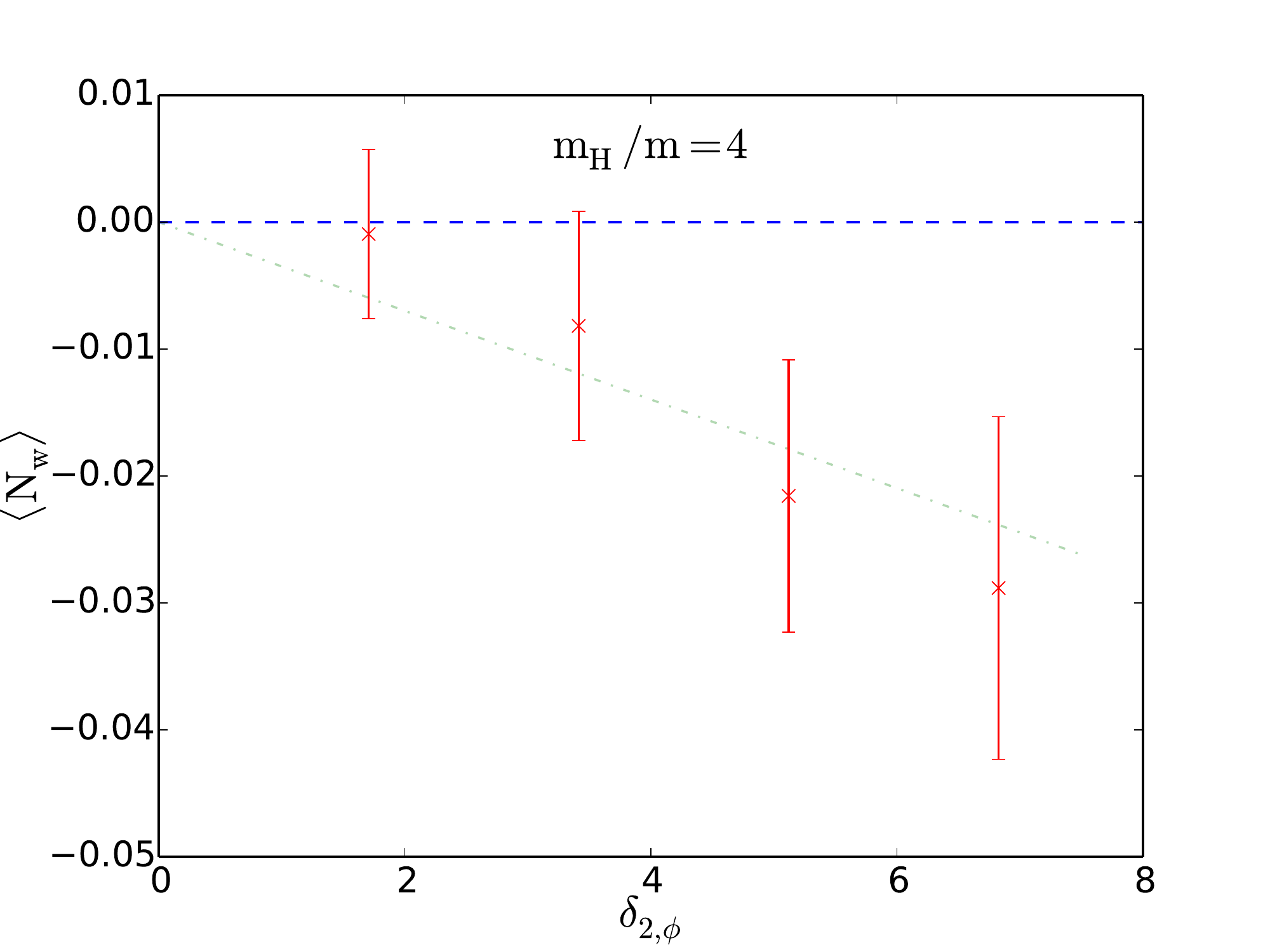}
\includegraphics[width=7cm,angle=0]{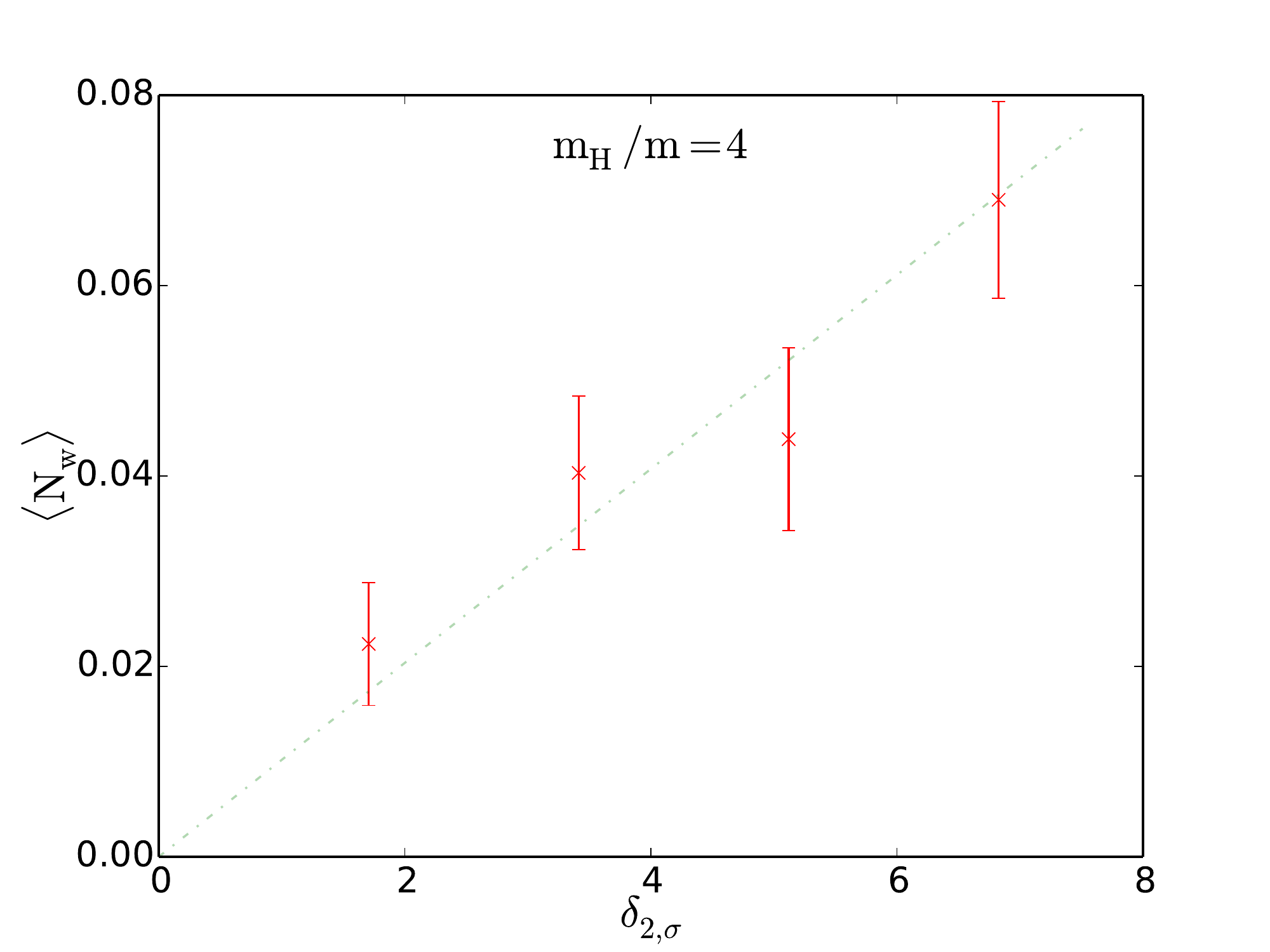}
\includegraphics[width=7cm,angle=0]{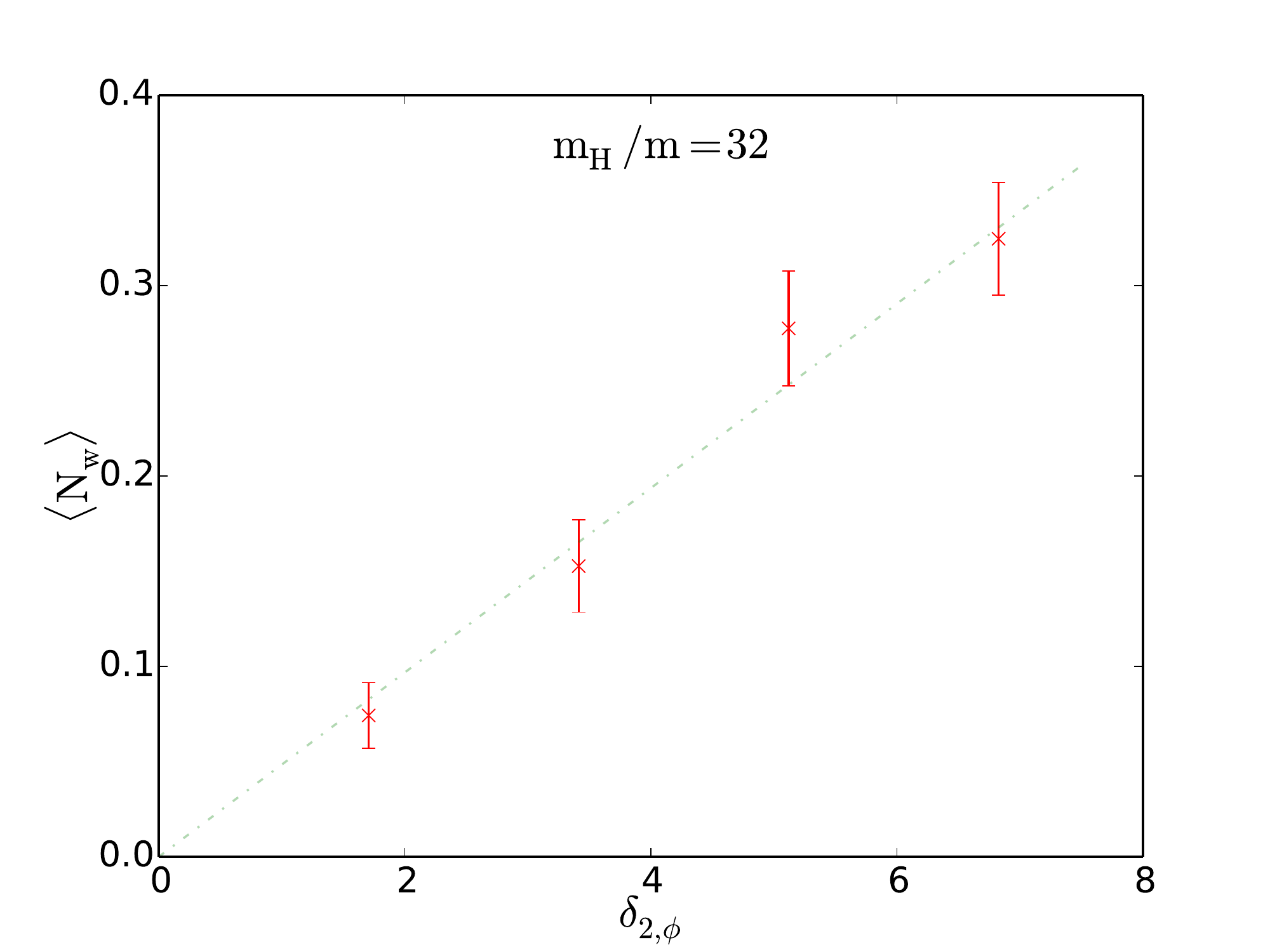}
\includegraphics[width=7cm,angle=0]{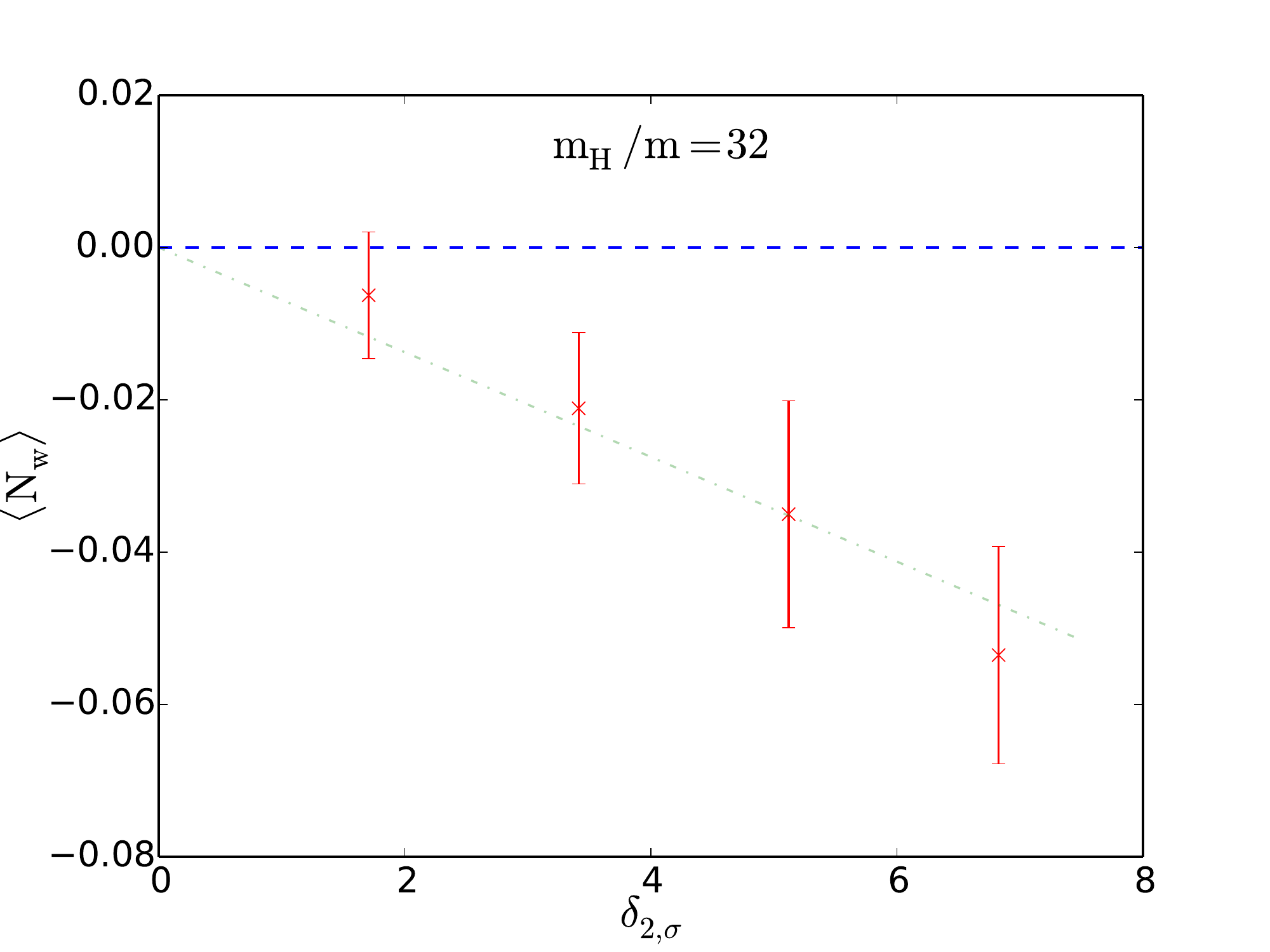}
\caption{The asymmetry for the type of primary CP-violation involving SU(2) gauge fields. Coupled to the Higgs field (left) and the singlet field (right). For fast (top) and slow (bottom) transitions.}
\label{fig:su2type}
\end{center}
\end{figure}

In Fig. \ref{fig:su2type} we show the final asymmetry in $N_{\rm w}$ for the two CP-violating terms involving the SU(2) gauge fields. In our terminology, they both represent a primary bias of Chern-Simons number. We show four separate cases, corresponding to fast (top) and slow (bottom) transitions, when the SU(2) field is coupled to the Higgs field (left) and when it is coupled to the $\sigma$ field (right). 

Concentrating first on the SU(2)-Higgs case, we notice is that the asymmetry is positive for slow quenches, and negative for fast quenches. For both quench times, the dependence on $\delta_{2,\phi}$ is linear, but with a much larger magnitude for the slow quench. We can fit the dependence with a 1-parameter form to find
\begin{eqnarray}
\langle N_{\rm w}(t)-N_{\rm w}(0) \rangle&= -(3.5\pm 0.7)\times10^{-3}\delta_{2,\phi}, \quad& \big(m_H/m=4,\,\textrm{SU(2)}-\phi\big)\\
&=(48\pm 2)\times10^{-3}\delta_{2,\phi}, \quad& \big(m_H/m=32,\,\textrm{SU(2)}-\phi\big).
\end{eqnarray}

When replacing the Higgs field by the $\sigma$ field, we anticipate that the prefactor of $W\tilde{W}$ ($\sigma$) is no longer (as) strongly correlated with the availability of Higgs zeros (in $\phi$).  But also, because $\sigma^2$ runs from finite positive to zero (so decreases in time), we expect the bias and hence the asymmetry to have the opposite overall sign. We indeed see this, and also that for a slow transition the asymmetry is reduced by a factor of about six compared to the Higgs-SU(2) term (for values of $\delta_{2,\sigma}$ similar to the  $\delta_{2,\phi}$ above). This is sensible, since the slow quench is specifically tuned to a maximum of Higgs zeros, rather than for instance where the CP-violating term is maximal. We see that for a fast transition, which does not optimize the availability of Higgs zeros, we get an asymmetry of the roughly the same magnitude, whether through Higgs-SU(2) or $\sigma$-SU(2). 

We may again fit with a linear relation, to find
\begin{eqnarray}
\label{eq:fastmax}
\langle N_{\rm w}(t)-N_{\rm w}(0) \rangle&= (10\pm 1)\times10^{-3}\delta_{2,\sigma}, \quad &\big(m_H/m=4,\,\textrm{SU(2)}-\sigma\big)\\
&=-(6.9\pm0.7)\times10^{-3}\delta_{2,\sigma}, \quad &\big(m_H/m=32,\,\textrm{SU(2)}-\sigma\big).
\end{eqnarray}

A rescaling of $\xi$ or $\sigma$ naively corresponds to changing $\delta_{2,\sigma}$, and so a priory, it is unclear why the asymmetries should match in magnitude for the same values of $\delta_{2,\sigma}$. But since $\xi\sigma_i=\mu=\sqrt{\lambda}v$ it is perhaps not so surprising that the order of magnitude is the same. What is remarkable is that the change in sign between fast and slow quenches remains. This really seems to be a generic feature of the process, distinguishing between fast and slow transition regimes.

Generalizing to a much broader class of $\sigma$ potentials, it is possible to engineer the $\sigma$ to increase from zero to a non-zero vev. From one vev to another. Or to/from a very large/small amplitudes. In each case, one will get a different asymmetry, which then again corresponds to a differently value of $\delta_{2,\sigma}$ and possibly a flipping of the sign, depending on when whether the $\sigma$ increases or decreases in magnitude.

\subsection{U(1)-type CP-violation}

\begin{figure}
\begin{center}
\includegraphics[width=7cm,angle=0]{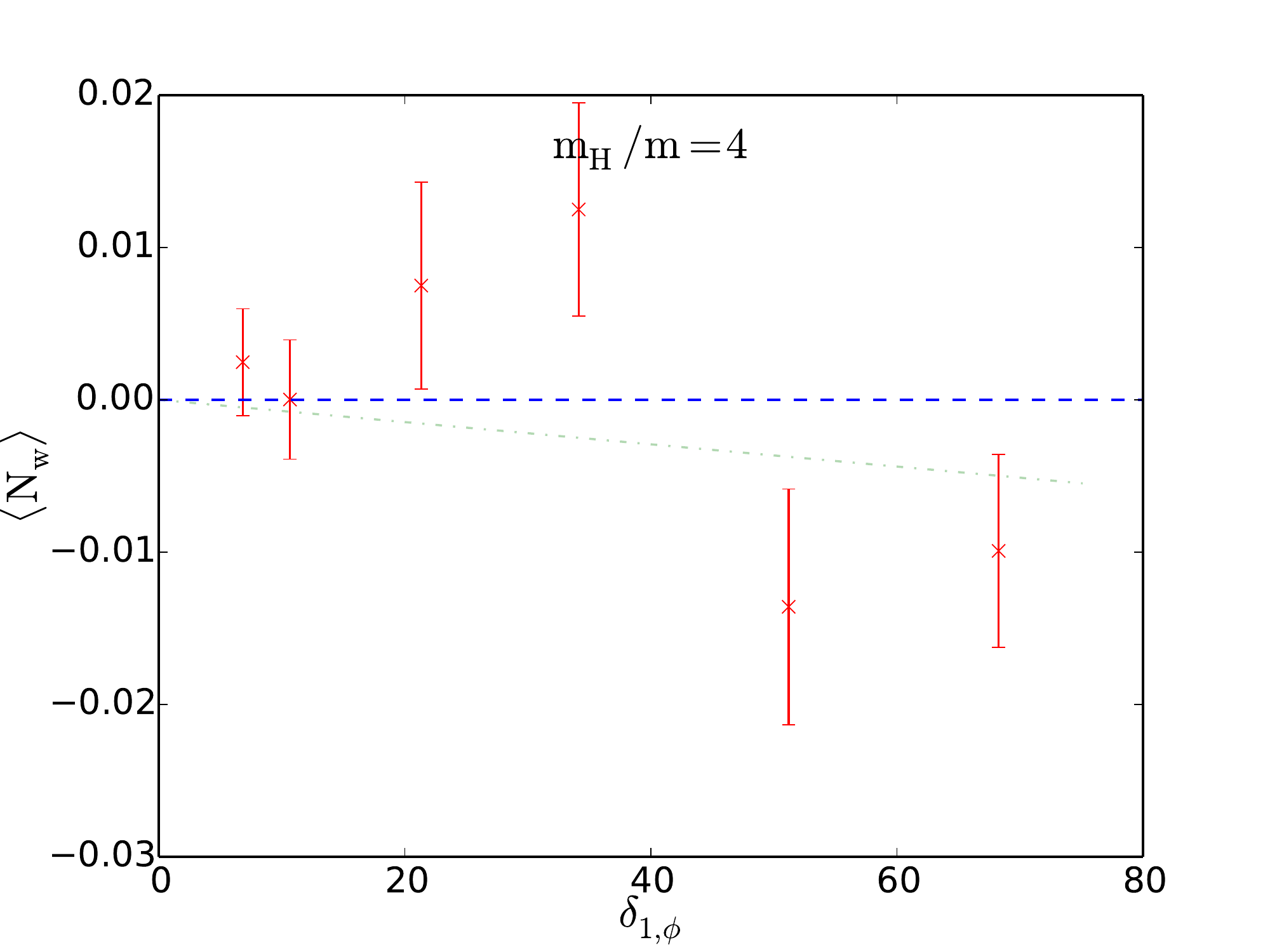}
\includegraphics[width=7cm,angle=0]{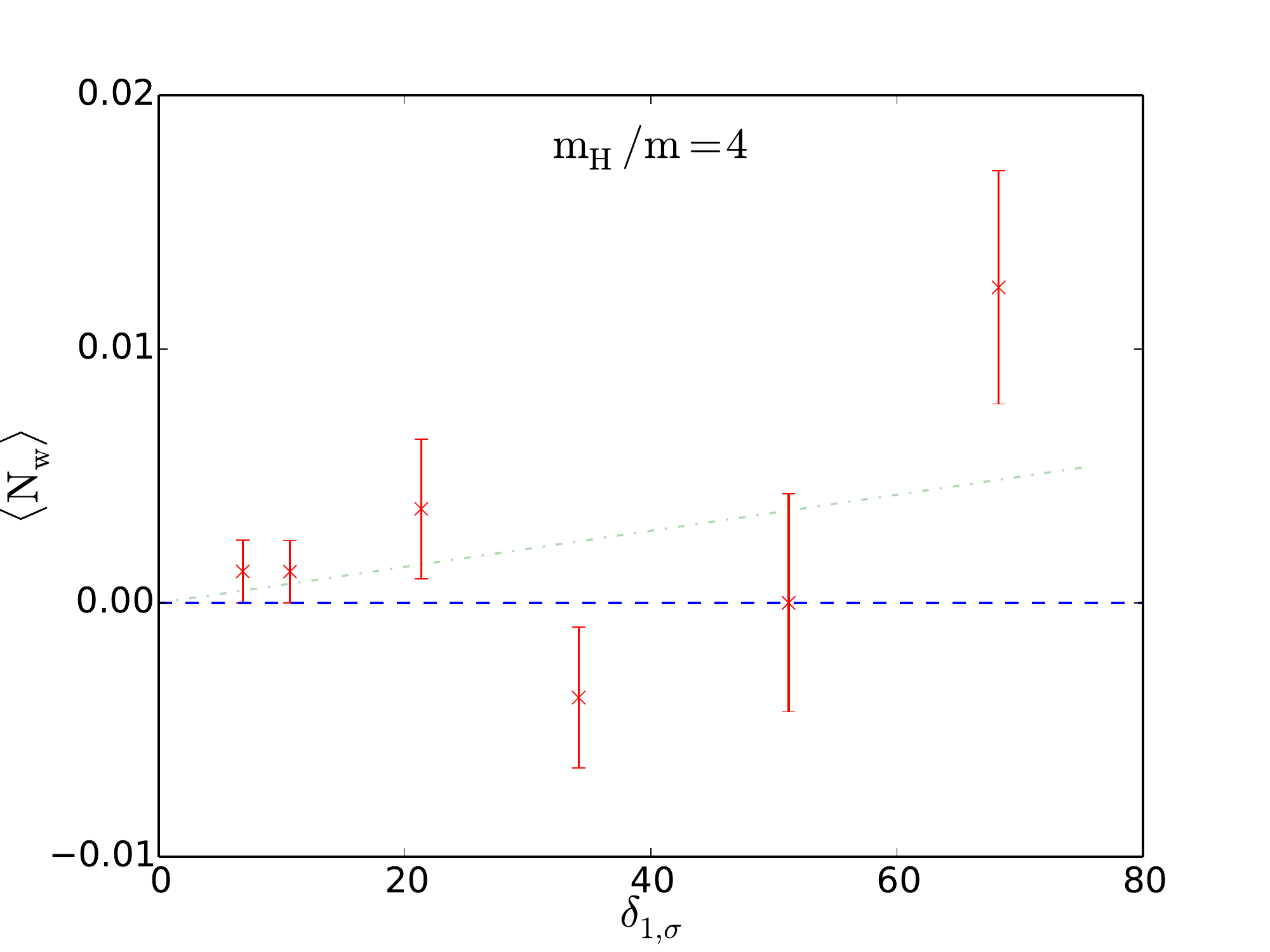}
\includegraphics[width=7cm,angle=0]{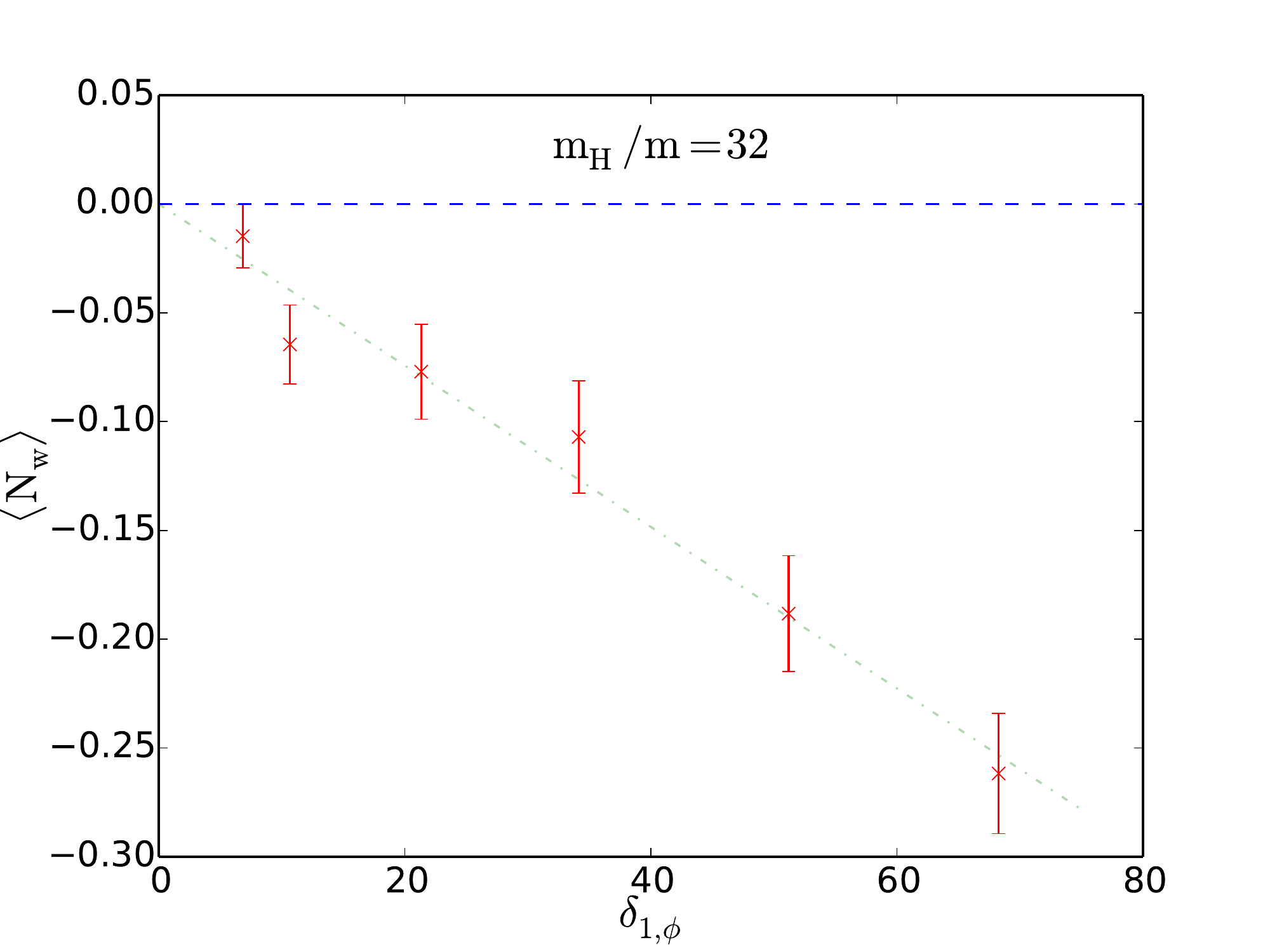}
\includegraphics[width=7cm,angle=0]{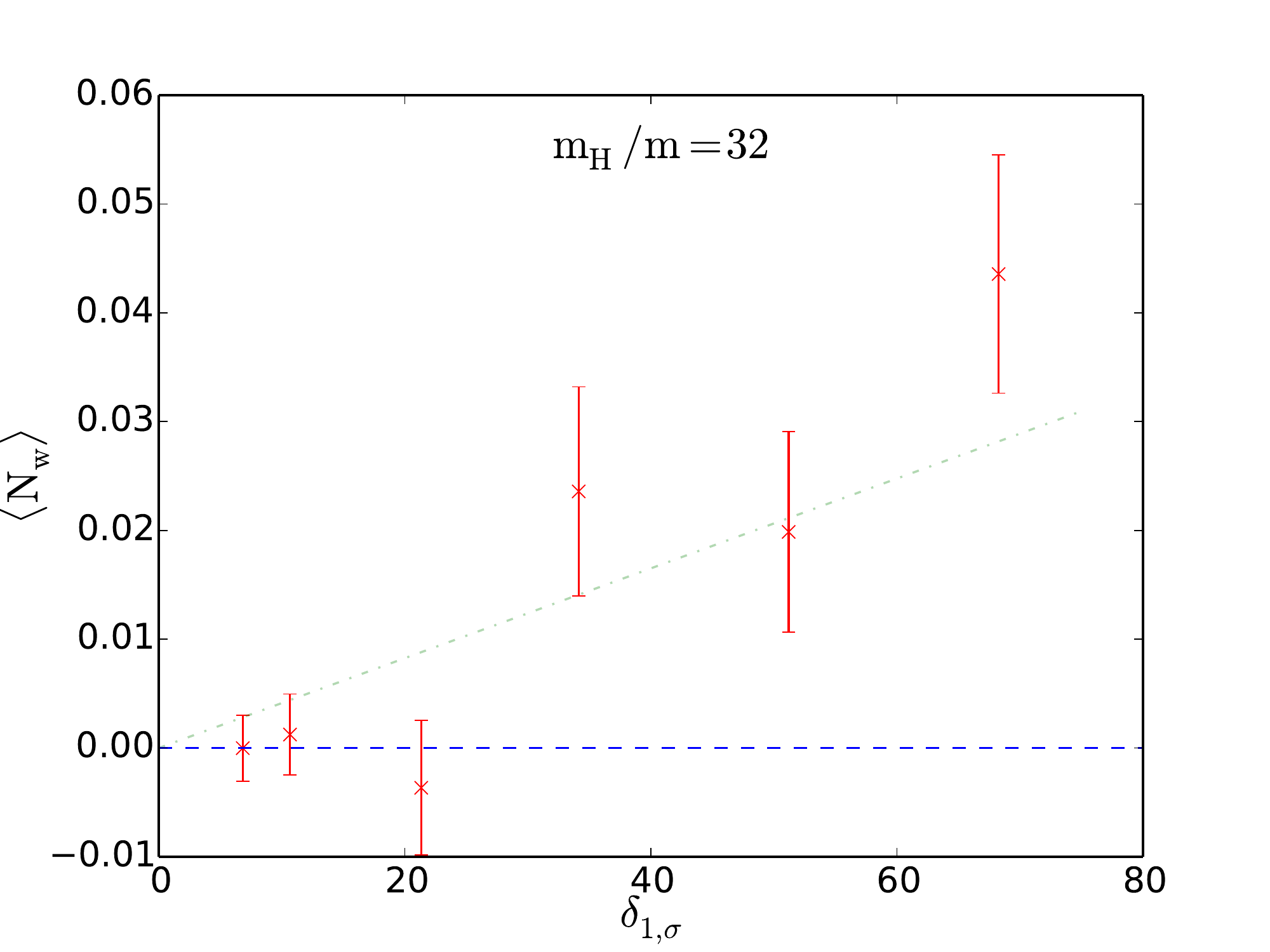}
\caption{The asymmetry for the type of CP-violation involving U(1) gauge fields. Coupled to the Higgs field (left) and the singlet field (right). For fast (top) and slow (bottom) transitions.}
\label{fig:u1type}
\end{center}
\end{figure}

In Fig. \ref{fig:u1type} we show a similar set of results, in the case where the gauge field in the CP-violating term is U(1) hypercharge. Now we have a situation where while the transition occurs, a U(1) gauge field is generated with non-zero Chern-Simons number, which then relaxes back to zero once the transition is over and thermalization completes. But while this Chern-Simons number is non-zero, the SU(2) gauge field and the Higgs field evolve in a (C)P-breaking background, leading to flips and a net asymmetry. That could in principle also relax back to zero, but because of the vacuum structure with high potential barriers in the low-temperature phase, leading to exponential suppression of Sphaleron transitions, once equilibrium is re-established the relaxation process takes longer than the age of the Universe. 

As for Fig. \ref{fig:su2type}, we show in the two lefthand panels the case where the bias is due to a coupling to the Higgs field. And in the right-hand panels, when we couple to the $\sigma$ field. The top panels are for a fast quench, $m_H/m=4$ and the bottom panels for a slow quench $m_H/m=32$. For each panel, we show the dependence on the strength of CP-violation.

We first note that the overall asymmetry of the U(1)-Higgs has the opposite sign to the SU(2)-Higgs system for positive $\delta_{i,j}$ (with our sign conventions, (\ref{eq:CP2p}), (\ref{eq:all3})). And the U(1)-$\sigma$ system has the opposite sign to the SU(2)-$\sigma$ system. Also, for the same values of $\delta_{i,j}$, the asymmetry in the U(1)-type systems is about an order of magnitude smaller than for the equivalent SU(2)-type terms of Fig. \ref{fig:su2type}. This is a question of normalization of the variables and prefactors of the CP-violating operator, but also indicates that the values of $B\tilde{B}$ are numerically smaller. 

For the fast quenches, both couplings to Higgs and $\sigma$ produce no statistically significant asymmetry. This may indicate that the asymmetry is in general very small for fast quenches, but most likely it is because $m_H/m=4$ happens to be where the dependence of the asymmetry on quench-time goes through zero on its way from positive to negative. The detailed quench speed dependence for the SU(2)-Higgs system was explored in \cite{Mou:2017atl}. For technical reasons to do with the lattice size, we are not able to reliably simulate even faster quenches (see again \cite{Mou:2017xbo}).

For slow quenches, we again find a clear asymmetry for both Higgs and $\sigma$-coupling, with a roughly linear dependence on the strength of CP-violation.  Just as for the SU(2)-type terms, the coupling to the Higgs field produces the largest asymmetry by a factor of 4-5. In terms of linear fits we find for the Higgs-U(1) term
\begin{eqnarray}
\langle N_{\rm w}(t)-N_{\rm w}(0)\rangle &= -(0.7\pm 1)\times 10^{-4}\delta_{1,\phi}, \quad& \big(m_H/m=4,\,\textrm{U(1)}-\phi\big)\\
&=-(37 \pm 2)\times 10^{-4}\delta_{1,\phi}, \quad &\big(m_H/m=32,\,\textrm{U(1)}-\phi\big).
\end{eqnarray}
and for the $\sigma$-U(1)
\begin{eqnarray}
\langle N_{\rm w}(t)-N_{\rm w}(0)\rangle &= (0.7 \pm0.5)\times10^{-4}\delta_{1,\sigma},\quad &\big(m_H/m=4,\,\textrm{U(1)}-\sigma\big)\\
&=(4 \pm 1)\times10^{-4}\delta_{1,\sigma}, \quad&\big(m_H/m=32,\,\textrm{U(1)}-\sigma\big).
\end{eqnarray}

\subsection{Adding up biases}
\label{sec:adding}

\begin{figure}
\begin{center}
\includegraphics[width=7cm,angle=0]{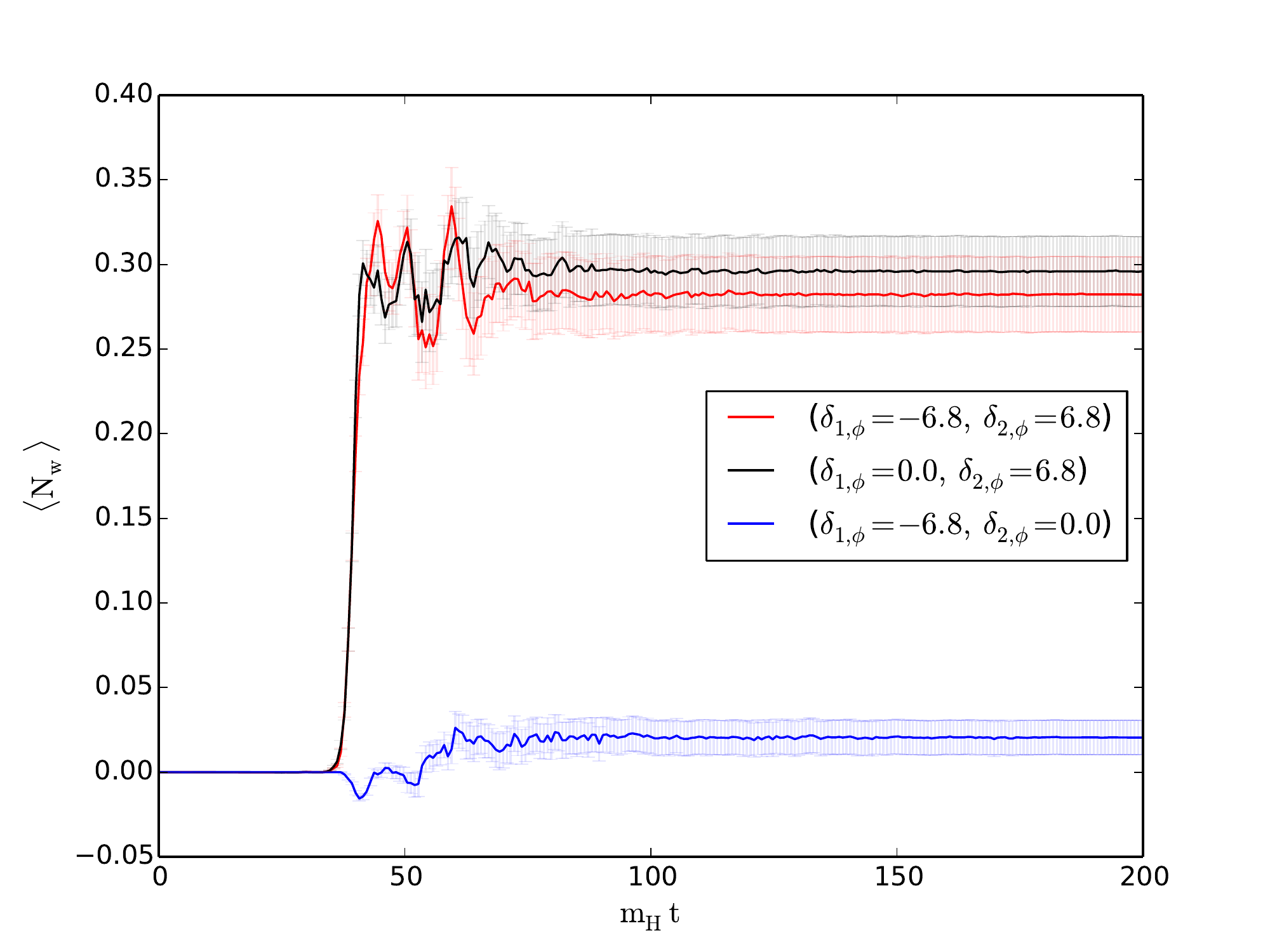}
\includegraphics[width=7cm,angle=0]{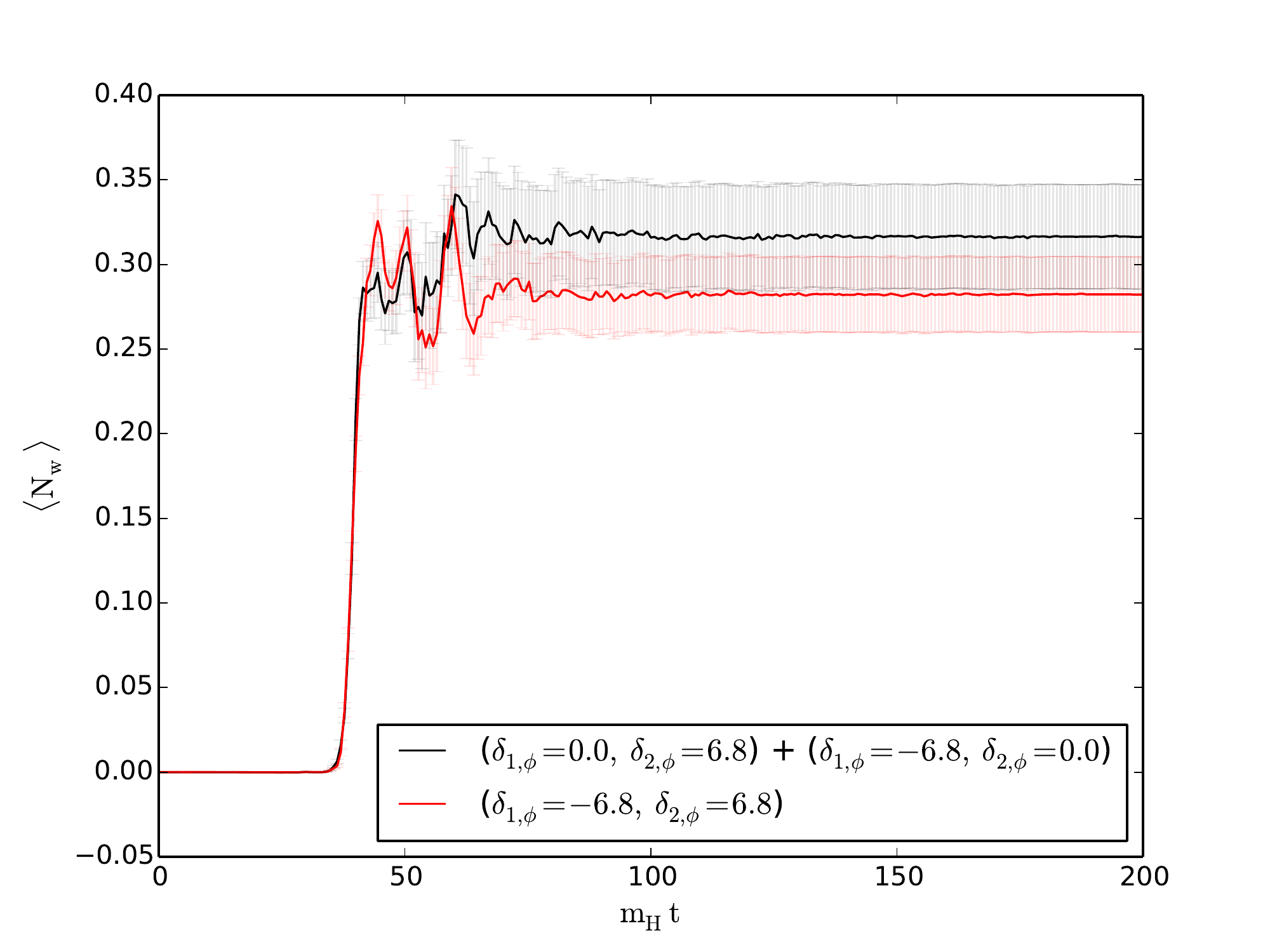}
\caption{The asymmetry from combining two CP-violating terms. Left: When only one source is on, and when two are on at the same time. Right: Comparing the sum of the two single-source asymmetries to the double-source asymmetry.  }
\label{fig:combine}
\end{center}
\end{figure}
Having computed the asymmetry from each of the four types of CP-violation, it is natural to ask what happens when two or more terms are active at the same time. This may of course be done in any number of different combinations, which different values of the four $\delta_{i,j}$. We will show one particular case here, namely
\begin{eqnarray}
S_{\rm 2+1,\phi} = \frac{3\delta_{2+1,\phi}}{m_{\rm w}^2} \phi^\dagger\phi\left(
\frac{g^2}{16\pi^2}\textrm{Tr } W^{\mu\nu}\tilde{W}_{\mu\nu}-\frac{(g')^2}{32\pi^2}B^{\mu\nu}\tilde{B}_{\mu\nu}
\right),
\end{eqnarray}
so that $\delta_{2,\phi}=-\delta_{1,\phi}=\delta_{2+1,\phi}=6.8$. By a similar argument to the one that led to (\ref{eq:bias}), we hence effectively bias the combination $N_{\rm cs,SU(2)}-N_{\rm cs,U(1)}$, which again through the anomaly equation is equal to the baryon number. We realise that this a very special choice, but it is just meant as one example of combining CP-violating terms. Since we have seen that in general, $\delta_{1,j}$ must be about an order or magnitude larger than $\delta_{2,j}$ to create the same size asymmetry, we expect the contribution from the SU(2) term to dominate.

In Fig. \ref{fig:combine}, we show the time-dependence of the Higgs winding number for three simulations, all at $m_H/m=32$. One run has only the Higgs-SU(2) term turned on (black line), another has only the Higgs-U(1) term turned on (blue line). And the third has both turned on simultaneously (red line). The bands around each curve correspond to one standard deviation on the average. In the left-hand plot, we show the individual three asymmetries, which grow and settle, with the U(1)-only asymmetry clearly the smallest, and the SU(2)-only asymmetry and SU(2)+U(1) asymmetry consistent within errors.

In the right-hand plot we compare the asymmetry from the combined run to the sum of the other two runs, according to $N_{\rm cs, SU(2)}-N_{\rm cs,U(1)}$. We see that the two agree within error bars. It seems that at least in this linear regime of the individual terms, combining multiple sources of CP-violation one may simply add up their individual contributions. No significant enhancements or suppressions arise. Although note that we chose a combination of terms precisely biasing the observable, we were intersted in. Whether for more generic combinations, competing effects create more complicated non-linear effects remains to be seen. Also, because the U(1) asymmetry is of the same order of magnitude as the statistical errors, we do not have the accuracy to make very strong statements on this point.

\subsection{Constant bias of SU(2) Chern-Simons number}
\label{sec:chemical}

\begin{figure}
\begin{center}
\includegraphics[width=7cm,angle=0]{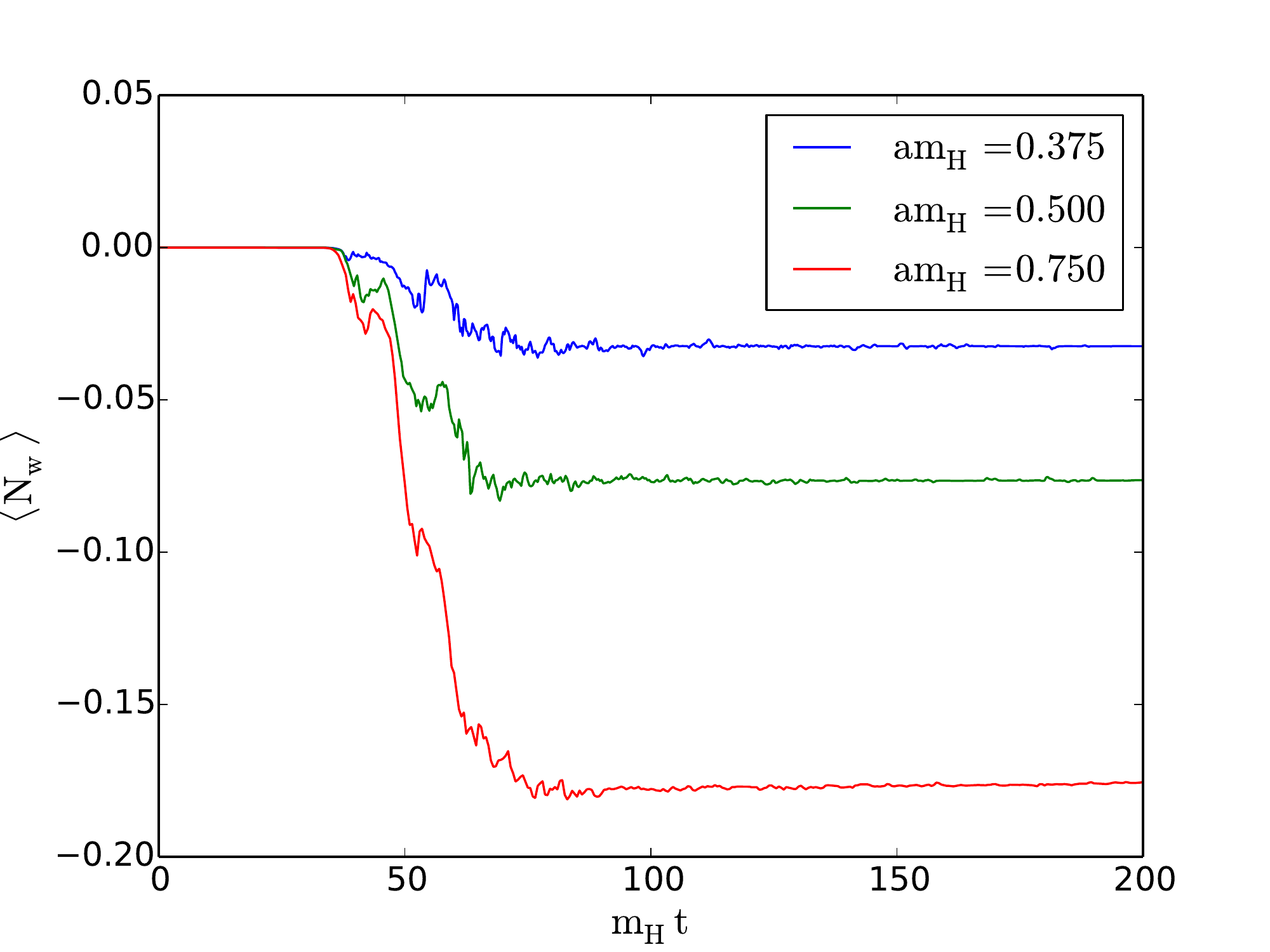}
\includegraphics[width=7cm,angle=0]{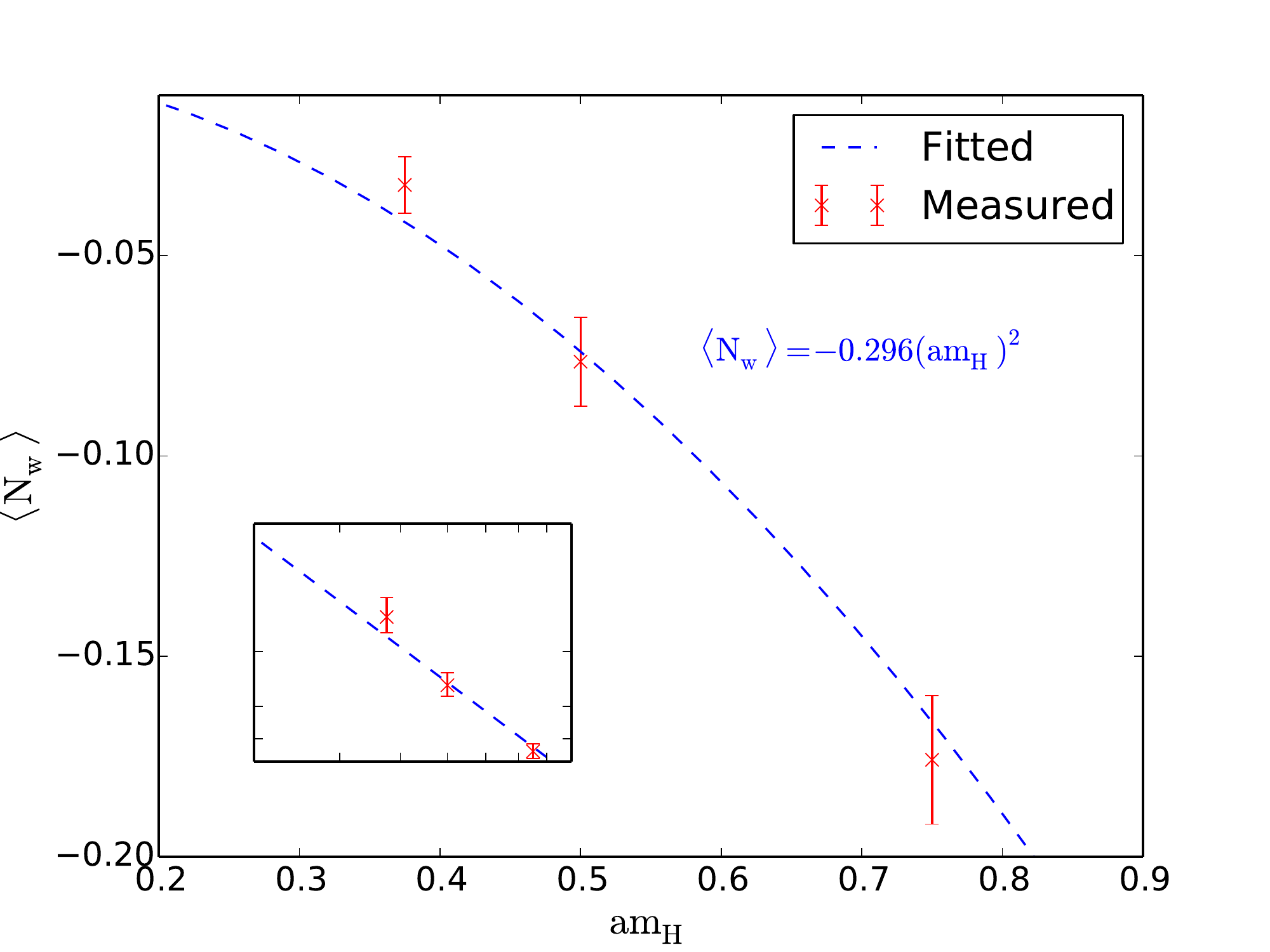}
\caption{The asymmetry from a constant bias for (lattice) $N_{\rm cs}$, for different lattice spacings with the same physical volume.}
\label{fig:adep}
\end{center}
\end{figure}

Since $W\tilde{W}$ is already responsible for breaking CP (through breaking P) in the simulations, one may imagine simply replacing the Higgs field by a constant, to get 
\begin{eqnarray}
S_{2}=\frac{3\delta_2g^2}{16\pi^2m_{\rm w}^2}\frac{v^2}{2}\textrm{Tr}\, W^{\mu\nu}\tilde{W}_{\mu\nu}=\frac{6\delta_2}{16\pi^2}\textrm{Tr}\, W^{\mu\nu}\tilde{W}_{\mu\nu}.
\end{eqnarray}
For a classical simulation, this should however not provide any asymmetry, since $W\tilde{W}$ is a total derivative, and so drops out of the equation of motion\footnote{At the quantum level, the story is different}. However the lattice implementation is not a total derivtaive at finite lattice spacing. Writing out the plaquette
\begin{eqnarray}
U_{x,\mu\nu}=U_{x,\mu}U_{x+\mu,\nu}U_{x+\nu,\mu}^\dagger U_{x,\nu}^\dagger= e^{-i a_\mu a_\nu F_{\mu\nu}^a\frac{\sigma^a}{2}+\mathcal{O}(a^4)}
\end{eqnarray}
This gives us, for small lattice spacing
\begin{eqnarray}
\textrm{Tr}\, W^{\mu\nu}\tilde{W}_{\mu\nu}\simeq \frac{1}{2}\epsilon^{\mu\nu\rho\sigma}\textrm{Tr}\, U_{x,\mu\nu}U_{x,\rho\sigma}=\frac{1}{2}\epsilon^{\mu\nu\rho\sigma}\textrm{Tr}\bigg[(1-ia_\mu a_\nu F_{\mu\nu}^a\frac{\sigma^a}{2}-\frac{a_\mu^2a_\nu^2}{2}F_{\mu\nu}^a\frac{\sigma^a}{2}F_{\mu\nu}^b\frac{\sigma^b}{2}+\cdots)\nonumber\\
\times (1-ia_\rho a_\sigma F_{\rho\sigma}^a\frac{\sigma^a}{2}-\frac{a_\rho^2a_\sigma^2}{2}F_{\rho\sigma}^a\frac{\sigma^a}{2}F_{\rho\sigma}^b\frac{\sigma^b}{2}+\cdots)\bigg],\nonumber\\
\end{eqnarray}
because of the antisymmetrization and the trace, what survives is
\begin{eqnarray}
\frac{1}{2}\epsilon^{\mu\nu\rho\sigma}\textrm{Tr} \,U_{x,\mu\nu}U_{x,\rho\sigma}=-\frac{1}{2}\epsilon^{\mu\nu\rho\sigma}\frac{a_\mu a_\nu a_\rho a_\sigma}{2} F^b_{\mu\nu}F_{\rho\sigma}^b+\mathcal{O}(a^6).
\end{eqnarray}
We find that to reduce lattice artefacts, it is necessary to symmetrize the plaquette as
\begin{eqnarray}
\bar{U}_{x,\mu\nu}=\frac{1}{4}\bigg(U_{x,\mu\nu}+U_{x,-\nu \mu}+U_{x,\nu -\mu}+U_{-\mu-\nu}
\bigg).
\end{eqnarray}
In any case the lattice term is not a total derivative, but has corrections of relative error expected to scale as $\mathcal{O}(a^2)$.

We may therefore expect CP-violating effects from this term, going to zero quadratically with the lattice spacing. 
In Fig. \ref{fig:adep} we compare simulations at equal physical volume, but lattice spacings of $am_H=0.375, 0.5, 0.75$. We use a quench time of $m_H\tau_q=32$ and $\delta_{2}=6.8$. We show the time histories of the Higgs winding number (left) and a fit to a purely quadratic dependence on lattice spacing (right). The fit is very convincing, confirming that the lattice artefacts contribute as expected. Also, the magnitude of the artefact contribution, although non-negligible, is subdominant relative to the total asymmetry once the dynamical Higgs field is reinstated. We note that all the above simulations were done at $am_H=0.375$, where the artefacts contribution is $\simeq -0.04$. As an estimate, this can be compared to the result for $S_{2,\phi}$ at the same $\delta_{2,\phi}=6.8$ of $0.33$, a systematic error of about 15\%. But it does teach us that using a larger lattice spacing could introduce systematic errors larger than the physical signal.

\section{Conclusion}
\label{sec:conclusion}

In a series of papers \cite{Mou:2017atl,Mou:2017zwe,Mou:2017xbo}, we have gradually relaxed simplifying assumptions on the dynamics and field content of simulations of Cold Electroweak Baryogenesis. The results show  that the main findings of the original work \cite{Tranberg:2003gi,Tranberg:2006ip,Tranberg:2006dg} are correct: A baryon asymmetry is produced in a tachyonic electroweak transition, as soon as CP-violation is present (primary or secondary). This asymmetry can be consistent with observations for reasonable values of the phenomenological dimensionless CP-violating parameters $\delta_{i,j}\simeq 10^{-5}$. The overall sign depends on the speed of quench, so that fast quenches, ``quench times" $m_H/m<4$, produce one sign (negative, in our conventions, for SU(2)-Higgs), and slower quenches produce the opposite sign. For very slow quenches $m_H/m > 60$, the asymmetry becomes very small. The replacement SU(2)$\rightarrow$U(1) flips the overall sign, and so does $\phi\rightarrow\sigma$.

The quantity of interest for observations is the baryon-to-photon ratio, and for the parameters used here, it is given by \cite{Mou:2017atl}
\begin{eqnarray}
\eta = \frac{n_{B}}{n_\gamma}= 8.55\times 10^{-4} \langle N_{\rm w}\rangle,
\end{eqnarray}
where $\langle N_{\rm w}\rangle$ refers to the specific simulations and lattice parameters described above. A sensible estimate is the to consider a fast quench for the SU(2)-Higgs term (\ref{eq:fastmax}), for which we find
\begin{eqnarray}
\eta = -9\times 10^{-6}\delta_{2,\phi}, 
\end{eqnarray}
and since the observed asymmetry is approximately $\eta=6\times 10^{-10}$, we require $\delta_{2,\phi}\simeq 7\times 10^{-5}$. Or 5 times smaller if we allow ourselves to tune to the optimal quench speed $m_H/m=32$.

This information can now be fed back to model building, where the largest caveat is how to engineer a cold symmetry breaking transition in the first place, while still triggering a fast enough quench. A few models exist on the market, where the $\sigma$ field may be identified with the inflaton \cite{vanTent:2004rc} or not \cite{Enqvist:2010fd} with the associated constraints from observations. And a more exotic scenario where the triggering is not due to a $\sigma$ but a supercooled phase transition \cite{Konstandin:2011ds,vonHarling:2017yew}. Much more work in this direction is required.

The second caveat is the origin of the CP-violation terms. The Standard Model does not provide large enough CP-violation \cite{Brauner:2011vb}, but the Two-Higgs Doublet Model (2HDM) might. If the Standard Model (or 2HDM or Standard Model+singlet) were a low-energy effective theory of something else, additional sources of CP-violation could be present from integrating out heavy degrees of freedom. 

This problem is not distinct from the lack of sufficient CP-violation in traditional (hot) Electroweak Baryogenesis. However, in the hot regime around a finite-temperature electroweak phase transition, temperature is around 160 GeV \cite{DOnofrio:2014rug}, which suppresses effective CP-violation. In the cold regime, we instead experience temperatures between near-zero (at the beginning) and up to 30-40 GeV after the transition. 

Ultimately, the true effective CP-violation will arise from integrating out heavy degrees of freedom in an out-of-equilibrium environment, a computation that is hard to do analytically.
In time, one would want to perform fully 3-family simulations of the whole SM + extensions with fermions, on large lattices with high statistics. Although the proof of method exists \cite{Saffin:2011kn}, the numerical effort is vast. 

For the moment, the highest priority seems to be to extend the set of viable and not too fine-tuned super-cooling and triggering mechanisms and scenarios, embedded in experimentally testable particle physics models. Since a fast triggering of Higgs symmetry breaking requires a sizeable coupling to whatever fundamental or composite BSM degree of freedom in whatever way, constraints from zero-temperature Higgs collider physics will be important. Standard portal couplings to what could be a Dark Sector could in turn connect baryogenesis to Darkmattergenesis, which could itself be based on a tachyonic transition or a more traditional first order phase transition. Getting all the numbers to match up (asymmetry, Dark Matter density, expansion of the Universe, evading direct detection, inflation) will likely require creativity in model building. 

\vspace{0.2cm}

{\noindent \bf Acknowledgements:} PMS is supported by STFC grant ST/L000393/1. AT and ZGM are supported by a UiS-ToppForsk grant. The numerical work was performed on the Abel supercomputing cluster af the Norwegian computing network Notur.

\end{document}